# A study of the magnetocaloric behavior of Dy-substituted $YMn_2O_5$ compounds

H. Chouaibi[1], S. Mansouri[2, 3], S. Aitjmal[4], M. Balli[4], O. Chdil[4], M. Abbasi Eskandari[3], S. H. Bukhari[5], P. Fournier[3]

[1]*Laboratoire de Physique Appliquée, Faculté des Sciences, Université de Sfax, B.P. 1171, 3000 Sfax, Tunisie.*

[2]*Institut National de la Recherche Scientifique, Énergie Matériaux Télécommunications, 1650, Boulevard Lionel-Boulet, Varennes, QC, J3X 1S2, Canada*

[3]*Institut Quantique et Regroupement québécois sur les matériaux de pointe, Département de physique, Université de Sherbrooke, J1K2R1, QC, Canada.*

[4] *AMEEC Team, LERMA, College of Engineering &Architecture, International University of Rabat, Parc Technopolis, Rocade de Rabat-Salé, 11100, Morocco.*

[5]*Department of Physics, Govt. College University Faisalabad, Layyah Campus, Layyah 31200, Pakistan.*

[*]***Correspondent author***: *chouaibichouaibi187@gmail.com*

**Abstract**

In this paper, we report on the magnetic and magnetocaloric features of Dy-substituted $YMn_2O_5$ ($Y_{1-x}Dy_xMn_2O_5$) with x =0.6, 0.8, and 1 series elaborated by sol-gel method. X-ray diffraction and Raman measurements attest well the high quality of our polycrystalline samples that crystallize in an orthorhombic structure with the *Pbam* space group. The Raman phonon frequencies were carried out and compared with the lattice dynamics calculations to identify the vibrational properties of all detected modes at room temperature. As expected, our magnetic study reveals that the magnetization was enhanced by the substitution of $Y^{3+}$ by $Dy^{3+}$. The Dy-substituted $YMn_2O_5$ sets the Néel transition [$T_N$ (Mn)] in the temperature range going from 40 to 45 K favoring the emergence of a transition at a very low temperature due to the long-range ordering of the $Dy^{3+}$ magnetic moments below 13K [$T_N$ (Dy)]. Dual peaks in the magnetic entropy change curve are also observed being in good agreement with magnetization data, which enlarges the range of application of these materials. On the other hand, a large magnetocaloric effect is observed close to 13 K which is mainly due the ordering of $Dy^{3+}$ magnetic moments. Also, the incommensurate antiferromagnetic transition of Mn magnetic moment taking place around 40 K affects slightly the entropy change. Our refrigerant capacity (RC) findings are higher compared to the average of RC for a, b, c axis of single crystal samples as $HoMn_2O_5$ and $TbMn_2O_5$, which confirms that our polycrystalline materials stand for promising magnetic refrigerant candidates that can be invested in space technology, hydrogen and helium liquefaction at cryogenic temperature.

## 1. Introduction

The fascinating physical properties of $RMn_2O_5$ multiferroic compounds such as $RMn_2O_5$ (R = rare earth) have attracted a wider interest from both engineers and scientists. These materials display a puzzling coupling between magnetic and ferroelectric orders, which enables the potential control of magnetization (M) an electrical polarization via external electric and magnetic fields (H), respectively [1]. This would open the way for their implementation in some promising applications, such as magnetoelectric sensors, multiferroic four-state memory, and microelectronics [2, 3]. The magnetoelectric $RMn_2O_5$ systems can also be used as a good magnetic refrigerant at cryogenic temperatures (T) [4, 5]. This is the case of $HoMn_2O_5$ and $TbMn_2O_5$ compounds that unveil both large conventional and rotating magnetocaloric effects (MCE) opening the way for their implementation in more compact, simplified, and efficient rotary cryomagnetic coolers [6, 7].

Magnetic refrigeration has been perceived by academia and industry as the next-generation cooling technique when compared to the standard vapor-compression cooling technology owing to its potential higher efficiency and ecofriendly character [8, 9]. Caloric effects are inherent in any material and they are defined as a change in entropy or temperature in response to an external excitation [10]. Usually, the magnetocaloric effect is characterized by the isothermal magnetic entropy change and the adiabatic temperature change upon varying external magnetic fields. If the temperature derivative of the magnetization is negative, a direct MCE can be observed, whereas, if it is positive, an inverse MCE occurs [11]. Our present work is devoted to the investigation of the magnetocaloric effect in the $Y_{1-x}Dy_xMn_2O_5$ polycrystalline samples. The family of $RMn_2O_5$ compounds (R = Dy, Y or Bi) [12] are typical examples of multiferroic materials at low temperatures. At room temperature, they crystallize in an orthorhombic structure with the space group *Pbam*. This latter consists of distorted $Mn^{4+}O_6$ octahedra sharing edges along the *c*-axis. The octahedra are linked by pairs of $Mn^{3+}O_5$ pyramids within the *ab*-plane [13].

On the other hand, the $RMn_2O_5$ multiferroics, display a complex and interesting phase diagrams with a cascade of magnetic and ferroelectric phase transitions below the Néel transition: $T_N$ = 40 – 45 K [14]. The $DyMn_2O_5$ compound exhibits an antiferromagnetic (AFM) order with an incommensurate magnetic modulation of the Mn magnetic moments at $T_{N1}$ = 43 K followed by a lock-in transition at $T_{C1}$ = 39 K to a commensurate ferroelectric phase. The $Mn^{3+}$ and $Mn^{4+}$ spins reorient further at $T_{N2}$ = 27 K and $T_{C2}$ = 18 K, respectively. The system re-enters into an incommensurate magnetic state at $T_{C3}$ = 6.5 K, which coincides with the ordering of $Dy^{3+}$ moments [15, 16]. Recently, the $HoMn_2O_5$ and $TbMn_2O_5$ single crystals have showed a giant rotating MCE at low temperatures. In the $HoMn_2O_5$ single crystal, the resulting rotating entropy change reaches a maximum value of 12.43 J/kg K at a constant magnetic field of 7 T [6]. Also, M. Balli et al. reported that under a constant relatively low magnetic field of 2 T, the rotating entropy change ($\Delta S_M$) is found to be 6.36 J/kg K for TbMn$_2$O$_5$ [7]. However, a few results about the MCE of RMn$_2$O$_5$ polycrystalline samples have been reported to exhibit large $\Delta S_M$ and refrigerant capacity (RC) in the literature compared to a single crystal. In addition, the synthesis of a single crystal is complicated compared to the elaboration of the polycrystalline sample. In the present paper, we investigate the magnetism and magnetocaloric effect in Y$_{1-x}$Dy$_x$Mn$_2$O$_5$ with x = 0.6, 0.8, and 1 compounds. As the Y is a non-magnetic ion, the substitution of Y$^{3+}$ by Dy$^{3+}$ would allow us to figure out whether the MCE in these materials is only related to the magnetic moments of Dy$^{3+}$ or also the exchange interactions between Mn$^{3+/4+}$ and Dy$^{3+}$ ions act on the thermal effects.

## 2. Experiment

Polycrystalline Y$_{1-x}$Dy$_x$Mn$_2$O$_5$ samples with (x = 0.6, 0.8, 1) were prepared by using sol–gel technique [17]. Stoichiometric amounts of Y$^{3+}$, Dy$^{3+}$ and Mn$^{3+/4+}$ nitrates were dissolved into distilled water. After it turned into a clean solution without any residues, citric acid was added. The obtained solution was stirred for 2 h along with heating at 80°C by using a magnetic agitator until a viscous gel was formed. The obtained gel was dried at 150°C for 2 h and finally the resulting black powder was sintered at 900°C for 6h. The crystalline structure as well as the quality of prepared samples were checked by X-ray diffraction (XRD) using a Bruker D8 Advance diffractometer with Cu-K$\alpha$ radiation of 2θ from 10° to 70° having a wavelength λ =1.5405Å. Magnetization data were collected in the temperature range 2 - 60 K under magnetic fields going from 0 to 7 T by using a **S**uperconducting **Q**uantum **I**nterference **D**evice (SQUID). The Raman spectra were recorded at room temperatures and obtained in the backscattering configuration using a He-Ne laser of 632.8 nm and a Labram-800 Raman spectrometer equipped with a microscope with an X-50 objective, an appropriate notch filter and a nitrogen-cooled CCD detector. The laser power was kept at ~0.8 mW to avoid local heating.

## 3. Results and Discussion

### 3.1. Structural properties

#### a. X-ray diffraction analysis

The obtained XRD patterns for Y$_{1-x}$Dy$_x$Mn$_2$O$_5$ (x=0.6, 0.8, and 1) were analyzed using the Rietveld refinement through the FullProf software [18] as shown in **Fig. 1**. The refined unit cell data and atomic positions are illustrated in **Table 1**. Lattice parameters display no noticeable shift in the position of diffraction peaks suggesting the same phase when increasing the Dy$^{3+}$ content. The entire XRD pattern unveils a well-crystallized orthorhombic phase with the space group *Pbam*. However, a slight increase in the lattice parameter was observed when going from Y$_{0.4}$Dy$_{0.6}$Mn$_2$O$_5$ to DyMn$_2$O$_5$. The obtained lattice parameters for all the here investigated samples are reported in **Table 1**. They are in good agreement with the already reported data [19, 20]. The numbers in the parentheses are the respective error bars. In such crystallographic Mn$^{3+/4+}$ ions are found in two different crystallographic sites [20] where $Mn^{4+}$ ions occupy the octahedral pyramidal site (4f) coordinated with six oxygen atoms. The $Mn^{3+}$ ions are located at a tetragonal pyramidal site (4h) with five oxygen atoms [20].

**Fig. 2** presents the variation of unit cell volume (V) as a function the $Dy^{3+}$ content (x). As can be seen, the increase of $Dy^{3+}$ concentration is associated with an increase in V. The increase of the lattice constants when rising $Dy^{3+}$ levels can be attributed to the difference in ionic radii of $Y^{3+}$ and $Dy^{3+}$ ions since. In fact, $Dy^{3+}$ ions (0.91Å) have greater ionic radii than those of $Y^{3+}$ (0.88Å) [21].

The average crystallite size of the synthesized samples was calculated by using Debye-Scherer's relation ($D_{SC}$) [22]:

$$D_{SC} = \frac{K\lambda}{\beta_{hkl} \cos \theta_{hkl}} \qquad (1)$$

where $D_{SC}$ is the average crystallite size (in nm), K is the shape factor (K = 0.9), λ is the wavelength of X-ray (1.5406 Å) Cu-Kα₁ radiation, β is the full width at half maximum of the diffraction peak, θ is the Bragg angle and (hkl) are the miller indices. It was found that the average crystallite size of $Y_{1-x}Dy_xMn_2O_5$ (YDMO) (x = 0.6, 0.8, and 1) increases with the content of Dy (**Table 1**).

The sizes of the crystallites were also determined based on the Williamson-Hall form ($D_{WH}$) given by [23, 24]:

$$\beta \cos(\theta) = \frac{K\lambda}{D_{WH}} + 4\varepsilon \sin(\theta) \qquad (2)$$

According to the curve $\beta \cos(\theta) = f(4 \sin(\theta))$ plotted in **Fig. 3**, the crystallite sizes $D_{WH}$ are estimated from a linear fitting. It is worthy to mention that the calculated sizes of crystallites $D_{WH}$ are higher than those calculated by Debye Scherer's formula ($D_{SC}$). These differences would be attributed to the effect of the enlargements produced by the strain Ɛ, which are equal to 0.023, 0.028 and 0.057% for x = 0.6, 0.8, and 1 samples, respectively. The obtained results are reported in **Table 1**. The nanocrystalline appearance of all the samples was checked by the values of particle size gained from Debye - Scherrer's and Williamson-Hall relations. The crystallite sizes for YDMO are found in nanometer dimension $D_{SC}$ (from 36 to 52 nm) and $D_{WH}$ (from 55 to 91 nm). In addition, it is obvious that the lattice strains are very small.

### b. Raman spectroscopy

**Fig. 4** shows the unpolarized typical Raman spectra of $Y_{1-x}Dy_xMn_2O_5$ (x = 0.6, 0.8, and 1) at room temperature. Typical Raman spectra associated with the orthorhombic $RMn_2O_5$ compounds are observed. The phonon frequencies are listed in the **Table 2**. The $A_g$ modes are expected to appear in the parallel *xx, yy,* and *zz* scattering configurations. The $B_{1g}$, $B_{2g}$, and $B_{3g}$ modes are allowed, in *xy, xz,* and *yz* configurations, respectively [25]. The assignment of the phonon symmetries refers to the previous works in Refs [26, 27]. The phonon frequencies of our polycrystaline sample $DyMn_2O_5$ match very well with those of the single crystals DyMn$_2$O$_5$ reported by García-Flores et al. [26]. Furthermore, the linewidths of the detected phonons are narrow attesting the high crystalline quality of our samples.

### 3.2. Magnetic properties

Magnetic measurements of Y$_{1-x}$Dy$_x$Mn$_2$O$_5$ (x = 0.6, 0.8, and 1) polycrystalline samples were performed using a SQUID with a maximum applied magnetic field H of 7T. Thermal variations of magnetization curves recorded at H = 1000Oe are displayed in **[Fig. 5(a, b, c)** respectively, for x = 0.6, 0.8, and 1**: in the left site: orange color]**. In the paramagnetic state (PM), the magnetization increases monotonically with the decrease in temperature. The curve of **dM/dT [Fig. 5(a', b', c')** for x = 0.6, 0.8, and 1, respectively**]** suggests that the temperature derivative of magnetization exhibits a crossover from negative to positive values, which is basically considered to be a characteristic of AFM transitions. **Fig. 5(a, b, c)** demonstrates that M (T) exhibits two successive transformations for all samples: an AFM - AFM transition at 13 K [$T_N$ (Dy)] [15, 34] and an AFM - PM transition at 43, 42 and 44 K [$T_N$ (Mn)] [28] for x = 0.6, x = 0.8 and x =1, respectively. These samples have several phase transitions being closer to those of the HoMn$_2$O$_5$ compound [29]. It is worthy to underline that the transition at 13 K refers to the long-range ordering of Dy$^{3+}$ moments and corresponds to the change of magnetic structure within the antiferromagnetically ordered state [30], which is consistent with the literature [31-32]. Mihailova et al. highlighted that the rare earth moments in HoMn$_2$O$_5$ and TbMn$_2$O$_5$ usually order below 15 K [31]. Besides, O. Prokhnenko asserted that below 15 K, the Dy$^{3+}$ magnetic moments in DyMnO$_3$ order also along the *b*-axis in a sinusoidal incommensurate state [32].

To further explore the nature of existing magnetic interactions, we plotted χ$^{-1}$ versus T **[Fig. 5 (a, b, c): in the right side: blue color]**. Based on the inflection points observed in χ$^{-1}$(T) curves, the $T_N$ (Mn) temperatures are found to be 43, 42 and 44 K for x = 0.6, x = 0.8 and x = 1, respectively. At high temperatures, χ$^{-1}$(T) curves follow the Curie–Weiss law given by [33]:

$$\chi = \frac{C}{T - \theta_{CW}} \qquad (3),$$

with the Curie constant (C) defined by the expression below [33]:

$$C = \frac{N_A \mu_B^2}{3k_B} \mu_{eff}^2 \qquad (4),$$

where $N_A$ = 6.022x10$^{23}$ mol$^{-1}$ is Avogadro's number, $\mu_B$ = 9.274x10$^{-21}$ emu is the Bohr magneton and $k_B$ = 1.38016x10$^{-16}$ erg/K is the Boltzmann constant. $\theta_{CW}$ is the Curie-Weiss temperature and $\mu_{eff}$ is the effective magnetic moment.

By fitting χ$^{-1}$(T) at high temperature to Curie-Weiss law, we have calculated the $\theta_{CW}$ and the experimental paramagnetic effective moments ($\mu_{eff}$):

$$\mu_{eff} = \sqrt{\frac{3k_B C}{N_A}} \qquad (5),$$

The obtained results are summarized in **Table 3**. The relatively high negative values of $\theta_{cw}$ suggest the predominance of strong antiferromagnetic (AFM) exchange interactions. $Y^{3+}$ ions have no magnetic moments since they do not have unpaired *d-* or *f-*electrons. Therefore, the magnetic state in $Y_{1-x}Dy_xMn_2O_5$ is determined uniquely by $Mn^{4+}$, $Mn^{3+}$ and $Dy^{3+}$ ions subsystems.

The theoretical effective magnetic moment is given by the following formula [34]:

$$\mu_{eff} = \sqrt{x\mu_{eff}^2(Dy^{3+}) + \mu_{eff}^2(Mn^{3+}) + \mu_{eff}^2(Mn^{4+})} \quad (6),$$

where the theoretical values of $\mu_{eff}(Mn^{3+})$, $\mu_{eff}(Mn^{4+})$ and $\mu_{eff}(Dy^{3+})$ are $4.90\mu_B$ and $3.87\mu_B$ and $10.63\ \mu_B$, respectively [35]. The theoretical values of $\mu_{eff}$ are also reported in **Table 3**. In this way, it is worthy to mention that the $Dy^{3+}$ substitution impacts the magnetic properties through enlarging Dy-Mn exchange interactions leading to an overall increase of the whole magnetic moment. As can be seen in **Table 3,** the theoretical values of $\mu_{eff}$ are slightly higher than the experimental ones. This can be probably ascribed to the existence of some ferromagnetic (FM) clusters in the PM state [36]. On the other hand, the obtained $\mu_{eff}$ further confirms that $Y^{3+}$ is non-magnetic in the $Y^{3+}$ oxidation state. Thus, the effective magnetic moment $\mu_{eff}$ mainly originates from the interaction between the magnetic moments of transition metal (Mn) elements with 3d-electrons and rare-earth ($Dy^{3+}$) elements with 4*f*-electrons [37, 38]. The values of calculated effective moments for all samples (**table 3**) are closer to those reported by Dutta et al. [34].

### 3.3. Magnetocaloric properties

To study the effect of the magnetic field on the magnetic state, magnetic isotherms are carried out in the temperature range between 2 and 60 K as shown in **Fig. 6 (a, b, c)**. We find that at low temperatures the magnetization increases rapidly with the applied magnetic field due to the metamagnetic transition, particularly for $DyMn_2O_5$ (x = 1). This increase becomes gradual without reaching saturation even at high magnetic fields, while the magnetization monotonically decreases with increasing temperature. The absence of saturation has been explained by the fact that both $Dy^{3+}$ and $Mn^{4+}$ elements carry magnetic moments [39], leading to a competition between the magnetic interactions at low temperatures [40]. This could be also explained by the strong coupling between the Mn sublattice magnetic moments [6, 7]

A linear extrapolation to zero magnetic fields of the magnetization data at higher applied magnetic fields provides the "experimental saturation magnetic moment" ($M_S^{exp}$) [39] at 5 K of 1.23 $\mu_B$ for 0.6 and 1.40 $\mu_B$ for x = 0.8 and 2.14 $\mu_B$ at 2 K for x=1.

On the other side, the "calculated saturation magnetic moment" per formula unit of the YDMO (x = 0.6, 0.8, and 1) compounds can be expressed as [39]:

$$M_S^{th}(\mu_B) = n(Dy^{3+}) \times M_S^{th}(Dy^{3+}) + n(Mn^{4+}) \times M_S^{th}(Mn^{4+}) \qquad (7),$$

where $M_S^{th}(Dy^{3+}) = g_J J \mu_B = 7.6\ \mu_B$, $M_S^{th}(Mn^{4+}) = g_s S \mu_B = 3\mu_B$ are the magnetic moments [39] (avoid the orbital contribution) $n(Dy^{3+})$ and as well as $n(Mn^{4+})$ are the contents of the $Dy^{3+}$ and $Mn^{4+}$ ions, respectively. A comparison between values $M_S^{exp}$ and the calculated ones $M_S^{th}(\mu_B)$ =10.5 6 $\mu_B$, 12.08 $\mu_B$, 13.6 $\mu_B$ for x = 0.6, 0.8, and 1, confirms the absence of full alignment of all spins, respectively. This difference also suggests the enhancement of the AFM frustration between $Dy^{3+}$ and $Mn^{4+}$ ions [39, 41].

The magnetocaloric properties of these compounds were studied by the common, indirect method using isothermal magnetization curves reported in **Fig. 6**. In this case, the magnetic entropy changes $\Delta S_M$ that can be calculated by using the well-known Maxwell equation given by [41]:

$$\Delta S_M(T, \Delta H) = \int_0^{\mu_0 H_{max}} \left(\frac{\partial M(\mu_0 H, T)}{\partial T}\right)_{\mu_0 H} dH \qquad (8),$$

where *M* is the magnetization, *H* is the applied magnetic field, and *T* is the temperature. The $\Delta S_M$ vs. T curves are plotted under several magnetic field changes in the temperature range covering the ordering zone of $Dy^{3+}$ and $Mn^{3+/4+}$ magnetic moments as shown in **Fig. 6 (a', b', c')**. The maximum values are superior to what is found in many other rare-earth-based materials [42]. One can note that the magnitude of $-\Delta S_M$ increases almost linearly with increasing the applied magnetic field and maximizes around the ordering point of $Dy^{3+}$ magnetic moments (13 K). Another peak that is related the ordering of Mn magnetic moments can be clearly seen close to 40 K for all the samples. Accordingly, the occurrence of two separate phase transitions [($T_N$(Dy), $T_N$(Mn)] gives rise to this observed dual peaks behavior, yielding significant entropy changes over a broad temperature range [41]. In the magnetic field change of 7 T, $-\Delta S_M^{max}$ reaches 5.99, 6.87, 9.49 J kg$^{-1}$K$^{-1}$ for x = 0.6, x = 0.8 and x = 1 respectively. It is worthy to mention that the magnetism and accordingly the magnetocaloric effect in $Y_{1-x}Dy_xMn_2O_5$ are strongly related to the spin correlation of the $Dy^{3+}$ and $Mn^{3+/4+}$ magnetic sublattices [6-7]. Also, the negative MCE is not seen in $Y_{1-x}Dy_xMn_2O_5$ (except for x = 1 at low temperature). The absence of negative MCE in compounds with x = 0.8 and 0.6 is more probably attributed to the weakening (destruction) of antiferromagnetic interactions under low magnetic fields as the concentration of magnetic $Dy^{3+}$ ions decreases. With increasing the doping content x, the spins associated with the $Dy^{3+}$-$Mn^{4+}$-$Dy^{3+}$ sub-lattice are progressively reinforced, and the rearrangement in the $Mn^{3+}$-$Mn^{4+}$-$Mn^{3+}$ block increases. Moreover, the spin coupling of $Dy^{3+}$-$Mn^{3+}$ ions becomes stronger than that of $Dy^{3+}$-$Mn^{4+}$ [43]. This behavior is mainly due to the complex arrangement of Mn moments in $RMn_2O_5$. In the latter, the magnetic moments $Mn^{3+}/Mn^{4+}$ are strongly coupled in AFM whatever the presence of the rare earth 4f magnetic moments [43], which makes the contribution of the moments $Mn^{3+}$ / $Mn^{4+}$ marginal to the total magnetization, being a common property of the multiferroic $RMn_2O_5$ [43, 7]. However, the *4f–3d* coupling between R and Mn ions can be very distinct for different members. Lee et al. reported that the Gd-Mn, Dy-Mn, and Ho-Mn couplings are strong while the Tb-Mn and Er-Mn couplings are relatively weak [44]. **Fig. 7(a, b)** depicts the $M_S^{exp}$ dependence of $-\Delta S_M^{max}$ for YDMO (x = 0.6, 0.8, and 1) samples. One can see that at $T_N$ (Dy) **Fig. 7(a)**, the curve is almost linear, which confirm that the increase of $-\Delta S_M^{max}$ is attributed to the high magnetic moment of the rare earth ($Dy^{3+}$) and we have a weak interaction between Mn and Dy ions. As well, the curve at $T_N$ (Mn) **Fig. 7(b)** is not linear, which assert that the increase of $-\Delta S_M^{max}$ is due to the coupling between Mn and Dy ions. Hence, the rise in the rate of Dy amplifies the magnetization and therefore increases the coupling Mn-Dy, which enlarges the $-\Delta S_M^{max}$.

**Fig.8 (a, b, c**: on the right hand for x = 0.6, 0.8, and 1, respectively**)** displays the applied magnetic field dependence of $-\Delta S_M^{max}$. One can see that the values of $-\Delta S_M^{max}$ are quite large, especially in the vicinity of $T_N$ (Dy) at H=7 T, which are considerab,ly higher than those reported in the single crystal $TbMn_2O_5$ where, the average of $-\Delta S_M^{max}$ for *a, b and c* axes are about 5.51 and 4.71J/kg. K at 5 T and 7 T respectively [45]. Also, our findings at 5 T and 7 T are superior to those reported by Balli et al. for the single crystal $HoMn_2O_5$ [46] whereby, the average of $-\Delta S_M^{max}$ for *a, b and c* axes are equal to 6.03 and 4.5 J/kg. K at 5 T and 7 T, respectively. As a result, the values of $-\Delta S_M^{max}$ for polycrystalline samples are higher than single crystal compounds. On the other hand, these values are significantly superior to the value of entropy change reported by Midya et al. [47] in $DyMnO_3$, that reaches 8 J/Kg. K for H = 7 T. Also, under 5 T around 13 K, $-\Delta S_M^{max}$ are equal to 4.02, 4.48, 6.27 for x=0.6, 0.8, 1 respectively, being higher than those reported by Hamdi et al. in $Dy_{0.5}(Sr_{1-x}Ca_x)_{0.5}MnO_3$, (3.5 J/kg K under 5T) [48]. Other compounds with similar working temperature range exhibit lower entropy changes when compared to the here investigated compounds, including for example, $Er_2Mn_2O_7$ [49], $DyCu_2$ [50] and $Eu_2CoMnO_6$ [51].

The refrigerant capacity (RC) is an important practical figure of merit that allows us to assess materials prior to their implementation in magnetic refrigeration. From a practical point of view, materials with a broad temperature range are convenient for the active magnetic refrigeration cycle [52]. In fact, RC depends not only on the maximum of -$\Delta S_M$ but also on the overall profile of -$\Delta S_M$ (T) curve. This parameter is obtained by the numerical integration of the area, under the -$\Delta S_M$ (T) curve. It represents the amount of heat that can be transferred between the hot and cold reservoirs in an ideal refrigerant cycle and can be expressed as follow [53]:

$$\mathbf{RC(\delta T, H)} = \int_{T_{hot}}^{T_{cold}} -\Delta SM\ (T)\ dT \qquad (10),$$

where $\delta T_{FWHM} = T_{hot} - T_{cold}$ is the full width at half maximum of -$\Delta S_M$ [54]. The applied magnetic field dependence of the found RC values using the above equations are reported in **Fig.8 (a, b, c:** on the left hand for x = 0.6, 0.8, and 1, respectively). As shown, RC increases notably in an almost linear profile with increasing magnetic field. These values are comparable to the reported values for numerous multiferroic materials with similar compositions [45, 46]. The RC value of YDMO reaches 251.84 J/kg under ΔH = 7 T and 153.09 J/kg under ΔH = 5 T at $T_N$ (Dy), being higher than those reported in the TbMn$_2$O$_5$ and HoMn$_2$O$_5$ single crystals [45, 46] where, the average of RC for a, b, c axis is equal to 178.5, 177 J/kg, under ΔH = 7 T **[Fig.9(c)]**, respectively. Thus, polycrystalline samples are better for magnetic refrigeration applications than single-crystal ones. Furthermore, the RC values at T$_N$ (Dy) of all samples at ΔH = 2T are larger than those reported in some Mn-based compounds, such as Ni$_2$MnGa (8.5 J/kg under ΔH = 2 T) [55]. Also, the RC values of 153.09 J/kg at 5 T are close to those reported in some rare-earth-based compounds at low temperatures such as DyCu$_2$ (166 J/kg at ΔH = 5 T) [50]. Thus, biphasic (two magnetic phase transitions) compounds would exhibit an enhanced RC in comparison to single-phased ones. These results unveil that YDMO with (x = 0.6, 0.8, 1) alloy exhibits an enhanced MCE. This enhancement refers to the high magnetic moment of Dy$^{3+}$ at T$_N$ (Dy), while at T$_N$ (Mn) it refers to the unusual magnetic ordering owing to the appearance of frustrated magnetic state of the Dy atoms located in between two antiferromagnetically coupled Mn layers. However, the shown magnetocaloric properties by YDMO polycrystalline samples remain lower when compared with RMnO$_3$ [6, 56-60] manganites and some orthovanadates such as RVO$_3$ [61-63] and RVO$_4$ [64], particularly along their easy axes.

**Conclusion**

In this work, the magnetic and magnetocaloric properties of Y$_{1-x}$Dy$_x$Mn$_2$O$_5$ (x =0.6, 0.8 and 1) compounds are explored. X-ray diffraction measurement indicated that the synthesized nanoparticles have a pure orthorhombic structure with Pbam space group. Raman spectroscopy was carried out and combined with lattice dynamics simulations to illustrate vibration properties of studied materials. When increasing the Dy content, an enhancement in the magnetization is detected. The Dy-substituted YMn$_2$O$_5$ sets the Néel temperature [T$_N$ (Mn)] in the range of 40 – 45 K and favors the appearance of a transition at a very low temperature due to the long-range ordering of the Dy$^{3+}$ moments below 13 K [T$_N$(Dy)]. We noticed dual peaks in the magnetic entropy change curve, in good agreement with the magnetization results confirming that the studied compounds demonstrate multi-magnetic phases behavior. The latter enlarges the region of application of these materials. Owing to the high magnetic moment of Dy$^{3+}$ ions, an important MCE appears below T$_N$ (Dy). Due to the antiferromagnetic ordering of Mn magnetic moments, MCE signatures around T$_N$ (Mn) are also observed. We assert an improvement of magnetocaloric effect with the increase in the Dy content. RC findings of our polycrystalline samples are higher compared to the average of RC for a, b, c axis values of HoMn$_2$O$_5$ and TbMn$_2$O$_5$ single crystals. The large refrigerant capacity shown by these compounds would make them interesting candidates for cryogenic application such as the liquefaction of hydrogen and helium.

**Acknowledgements**

The authors would like to acknowledge the financial support of the research project by the Tunisian Ministry of Higher Education and Scientific Research, the University of Sherbrooke (Canada), and the International University of Rabat.

**Figure captions**

**Figure 1:** The X-ray diffractogram with the Rietveld refinement of $Y_{1-x}Dy_xMn_2O_5$ (YDMO) (x=0.6, 0.8 and 1) compounds. The black circles in the XRD patterns correspond to the experimental data, the continuous red line represent the fitted curve, the vertical blue curves are the Bragg positions, and the difference between experimental and fitted patterns is exposed as a solid green line at the underside.

**Figure 2:** The cell volume V as a function of x (x=0.6, 0.8 and 1).

**Figure 3:** Williamson-Hall analysis curve for x=0.6, 0.8 and 1 samples.

**Figure 4:** The unpolarized Raman spectra for x=0.6, 0.8 and 1 compounds at room temperature obtained using a 632.8 nm excitation wavelength and at ~0.8 mW of laser power.

**Figure 5:** Temperature dependence of the magnetization under a magnetic field of 1000 Oe for x=0.6 **(a)**, x=0.8 **(b)** and x=1 **(c)** samples**:** (left column) **orange color]**, thermal variation of the inverse of susceptibility for x=0.6 **(a)**, x=0.8 **(b)** and x=1 **(c)** samples**:** (right column) **blue color]** and thermal magnetization derivative dM/dT for x=0.6 **(a')**, x=0.8 **(b')** and x=1 **(c')** alloys.

**Figure 6:** Isotherms M (H) curves for x=0.6 **(a)**, x=0.8 **(b)** and x=1 **(c)** samples and $-\Delta S_M$ (T) curves for x=0.6 **(a')**, x=0.8 **(b')** and x=1 **(c')** compounds.

**Figure 7:** $M_S^{exp}$ dependence of $-\Delta S_M^{max}$ at $T_N$ (Dy) **(a)** and at $T_N$ (Mn) **(b)** for x=0.6, 0.8 and 1 alloys.

**Figure 8:** The magnetic field dependence of the magnetic entropy change $-\Delta S_M^{max}$ **(on the right hand)** and the refrigerant capacity **(on the left hand)** for x=0.6 **(a)**, x=0.8 **(b)** and x=1 **(c)** alloys.

**Figure 9:** Refrigerant capacity shown by polycrystalline $DyMn_2O_5$ and single crystals $TbMn_2O_5$ and $HoMn_2O_5$ in the magnetic field change of 7T.

**Table captions**

**Table 1:** Structural parameters at room temperature for YDMO (x=0.6, 0.8 and 1) compounds.

**Table 2:** The phonon frequencies of the unpolarized Raman spectra for YDMO (x=0.6, 0.8, and 1).

**Table 3:** Magnetic properties of YDMO (x=0.6, 0.8 and 1) compounds.

Table 1

| Samples | $Y_{0.4}Dy_{0.6}Mn_2O_5$ | $Y_{0.2}Dy_{0.8}Mn_2O_5$ | $DyMn_2O_5$ |
|---|---|---|---|
| Space group | Pbam | Pbam | Pbam |
| a (Å) | 7.295(20) | 7.300(16) | 7.305(19) |

| | | | | |
|---|---|---|---|---|
| **b(Å)** | | 8.501(45) | 8.505(02) | 8.508(35) |
| **c(Å)** | | 5.681(29) | 5.683(22) | 5.683(70) |
| **V(Å³)** | | 352.352 | 352.859 | 353.277 |
| **$D_{SC}$ (nm)** | | 36 | 45 | 52 |
| **$D_{WH}$(nm)** | | 55 | 58 | 91 |
| **Ɛ (%)** | | 0.023 | 0.028 | 0.057 |
| **Y/Dy** | x | 0.359(76) | 0.360(54) | 0.359(88) |
| | y | 0.327(03) | 0.329(40) | 0.328(56) |
| | z | 0.5 | 0.5 | 0.5 |
| | Occupation | 0.257(12)/0.2 | 0.4/0.1 | 0.416(95) |
| **$Mn_1$** | x | 0.088(41) | 0.088(73) | 0.087(89) |
| | y | 0.15 | 0.150(26) | 0.15 |
| | z | 0 | 0 | 0 |
| | Occupation | 0.493(84) | 0.5 | 0.449(99) |
| **$Mn_2$** | x | 0 | 0 | 0 |
| | y | 0.5 | 0.5 | 0.5 |
| | z | 0.250(40) | 0.244(76) | 0.250(18) |
| | Occupation | 0.5 | 0.5 | 0.462(93) |
| **$O_1$** | x | 0.111(75) | 0.111(65) | 0.113(40) |
| | y | 0.293(97) | 0.296(31) | 0.295(06) |
| | z | 0.252(26) | 0.247(65) | 0.250(84) |
| | Occupation | 1 | 1 | 0.896(90) |
| **$O_2$** | x | 0.347(93) | 0.336(44) | 0.350(21) |
| | y | 0.052(20) | 0.056(71) | 0.057(78) |
| | z | 0.5 | 0.5 | 0.5 |
| | Occupation | 0.5 | 0.5 | 0.418(66) |
| **$O_3$** | x | 0.347(90) | 0.342(86) | 0.334(86) |
| | y | 0.065(29) | 0.067(15) | 0.063(21) |
| | z | 0 | 0 | 0 |
| | Occupation | 0.5 | 0.5 | 0.516(56) |
| **$O_4$** | x | 0 | 0 | 0 |
| | y | 0 | 0 | 0 |
| | z | 0.221(18) | 0.222(84) | 0.246(71) |
| | Occupation | 0.5 | 0.5 | 0.496(84) |
| **$R_p$ (%)** | | 22.6 | 22.3 | 18.4 |

| | | 25.8 | 25.7 | 21.4 |
|---|---|---|---|---|
| $R_{wp}(\%)$ | | | | |
| $\chi^2$ | | 3.46 | 3.39 | 1.18 |

**Table 2**

| Samples / Phonon symmetries | $Y_{0.4}Dy_{0.6}Mn_2O_5$ (cm$^{-1}$) | $Y_{0.2}Dy_{0.8}Mn_2O_5$ (cm$^{-1}$) | X=1: $DyMn_2O_5$ (our work) (cm$^{-1}$) | $DyMn_2O_5$ [27] (cm$^{-1}$) |
|---|---|---|---|---|
| | | | | |
| $A_g$ | 216 | 216 | 217 | 215 |
| $B_{1g}$ | 241 | 239 | 237 | 235 |
| $A_g$ | 273 | 273 | 275* | --- |
| $A_g$ | 349 | 349 | 350 | 350 |
| $B_{2g}$ | 457 | 457 | 457 | 460 |
| $A_g$ | 501 | 502 | 502 | 500 |
| $B_{1g}$ | 542 | 542 | 541 | 540 |
| $B_{3g}$ | 577 | 577 | 577* | -- |
| $A_g$ | 625 | 625 | 625 | 625 |
| $B_{1g}$ | 681 | 678 | 676 | 675 |
| $A_g$ | 695 | 695 | 695 | 695 |

**Table 3**

| Samples | $Y_{0.4}Dy_{0.6}Mn_2O_5$ | $Y_{0.2}Dy_{0.8}Mn_2O_5$ | $DyMn_2O_5$ |
|---|---|---|---|
| $T_N(Dy)(K)$ | 13 | 13 | 13 |
| $T_N(Mn)(K)$ | 43 | 42 | 44 |
| C (emu. K. mol$^{-1}$) | 12.52 | 14.28 | 16.66 |
| $\theta_{cw}$ (K) | -31 | -25 | -20 |
| $\mu_{eff}(\mu_B)^{exp}$ | 10 | 10.86 | 11.54 |
| $\mu_{eff}(\mu_B)^{theo}$ | 10.33 | 11.37 | 12.32 |

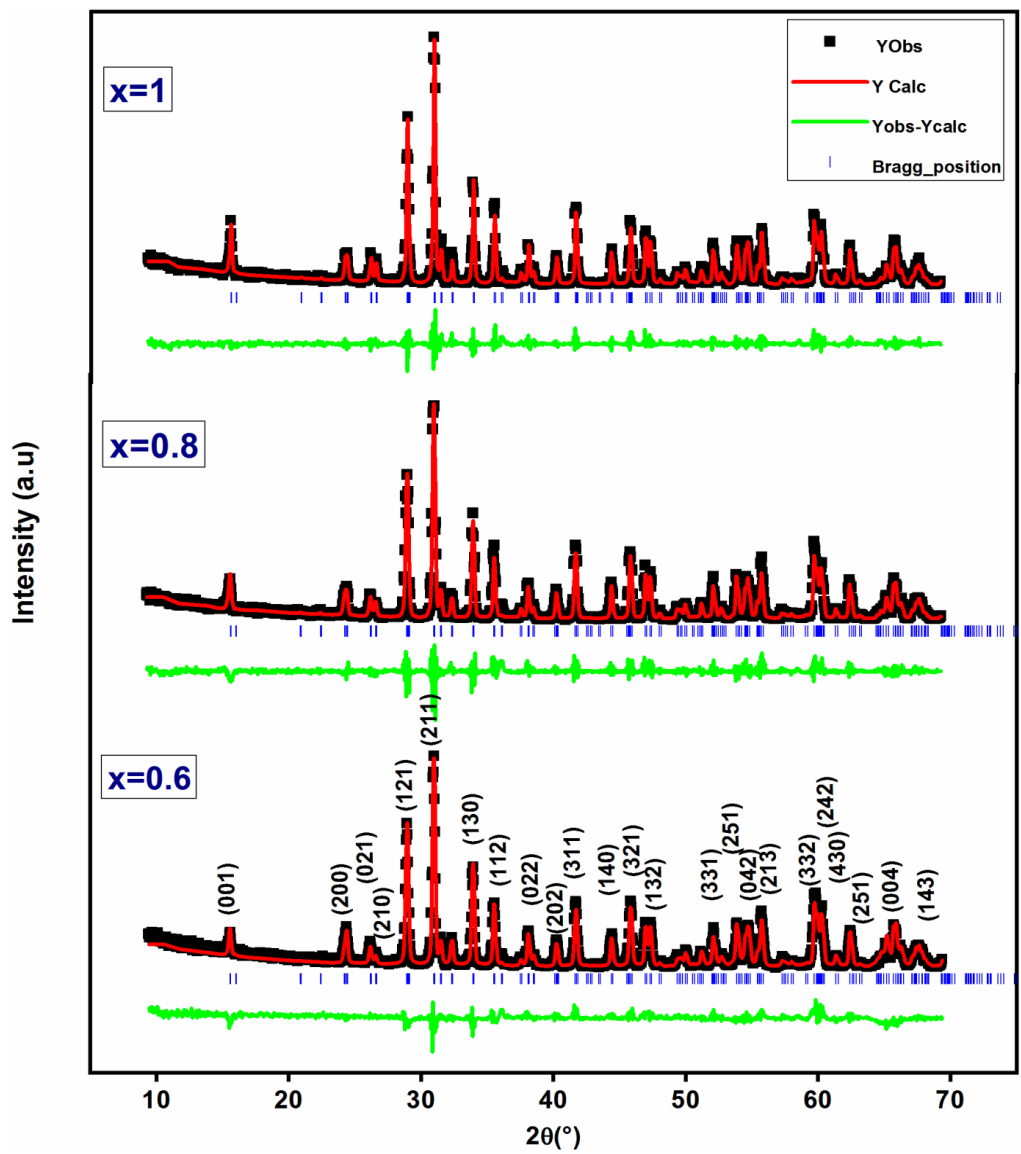

**Fig. 1**

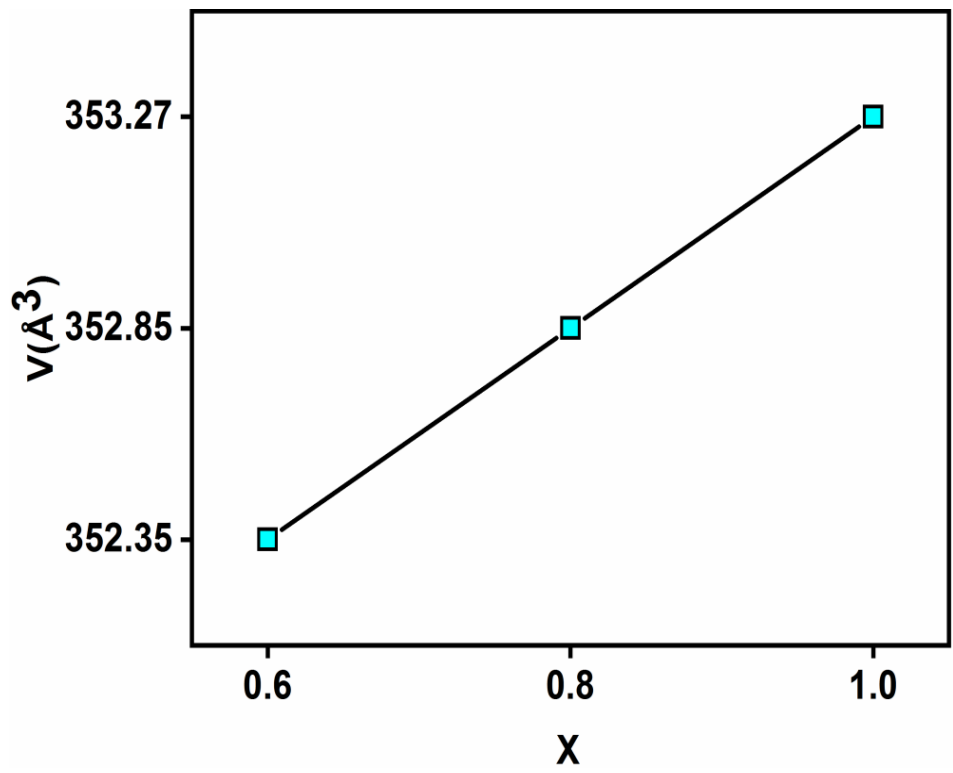

**Fig. 2**

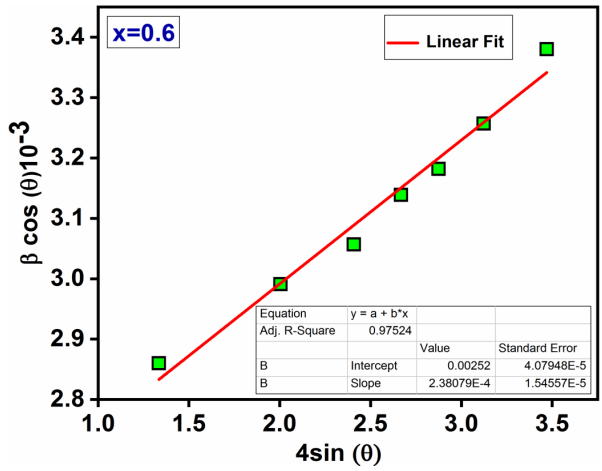

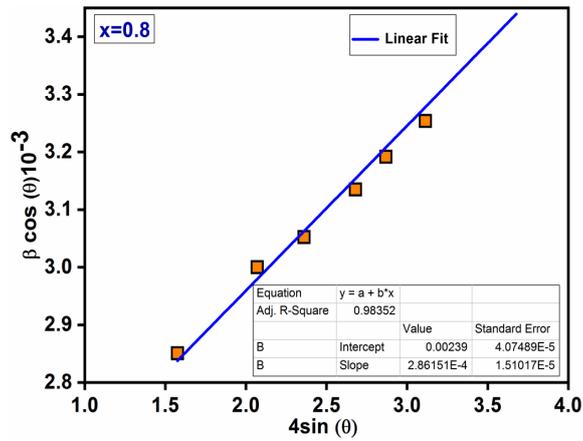

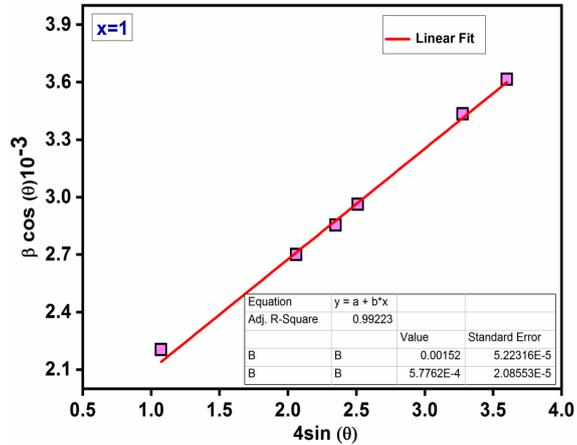

**Fig. 3**

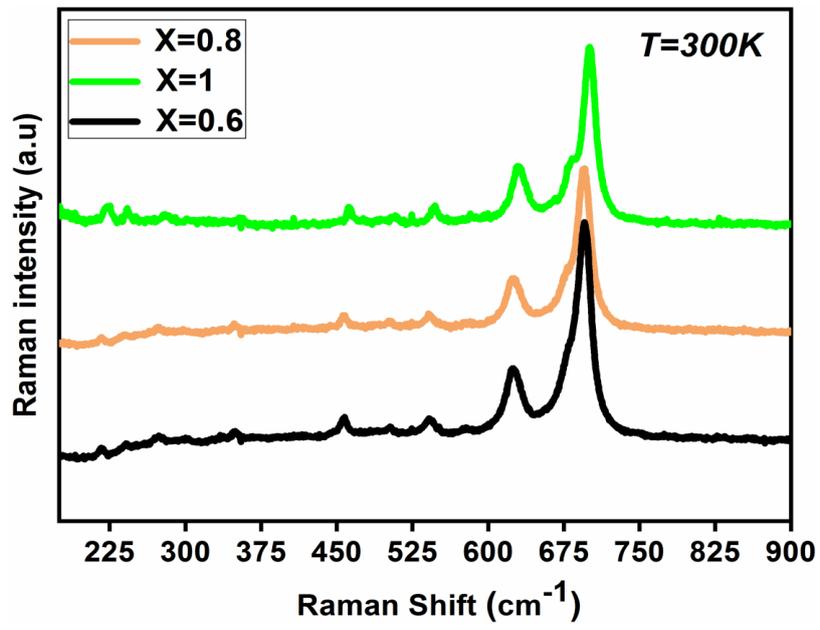

Fig.4

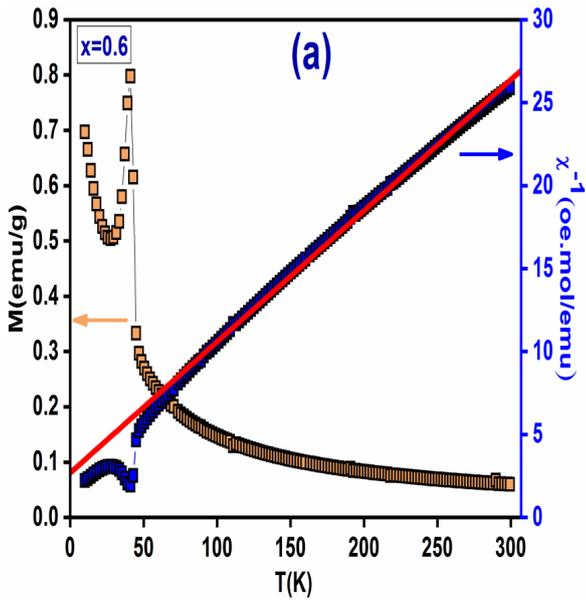
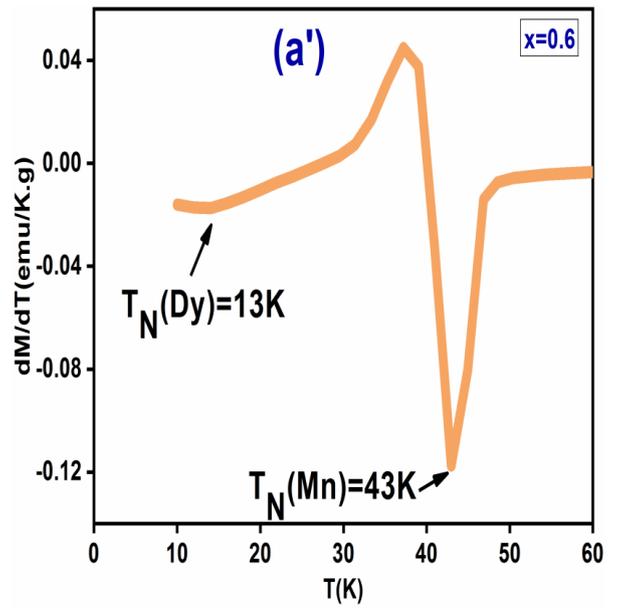
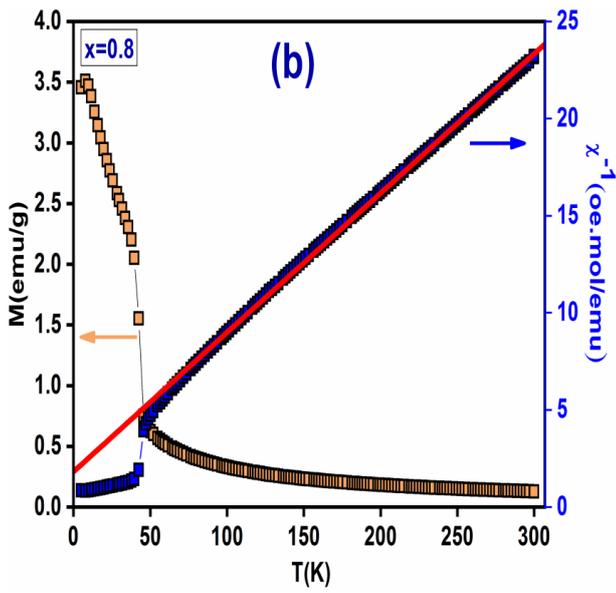
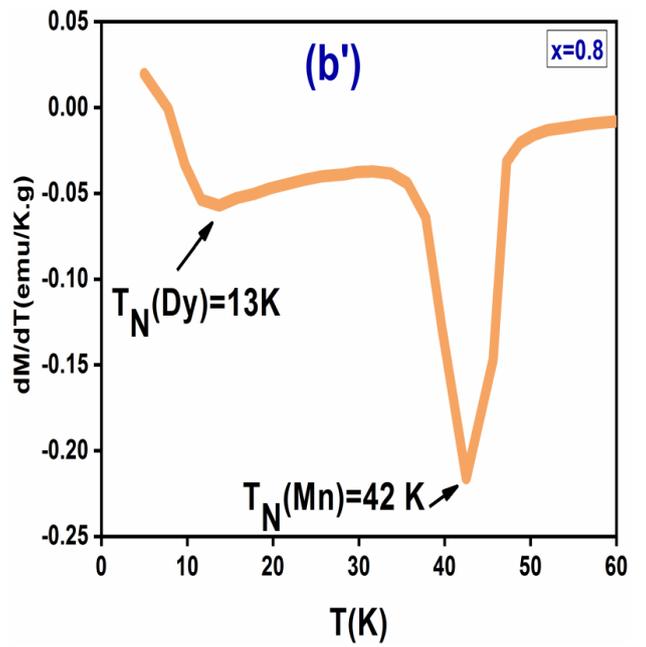

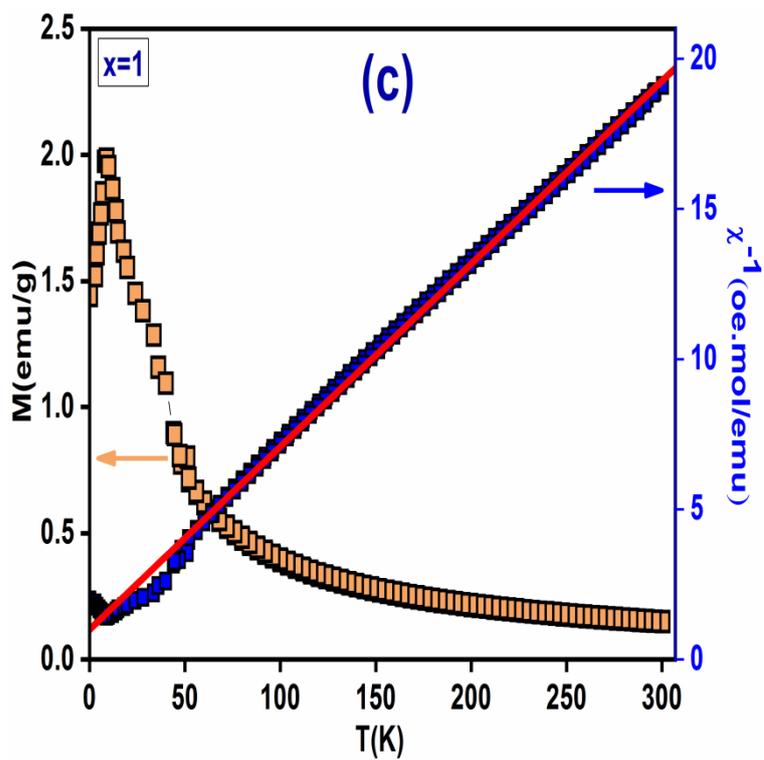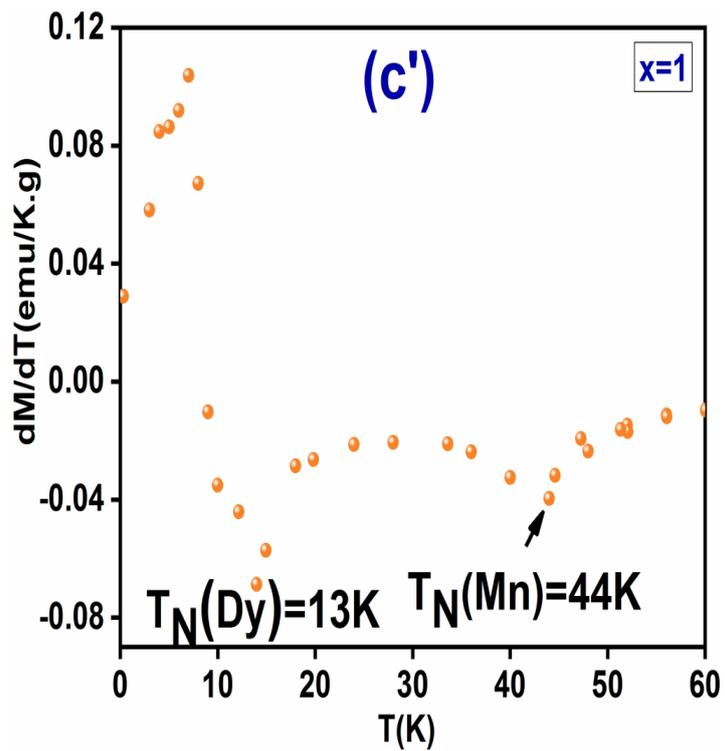

Fig. 5

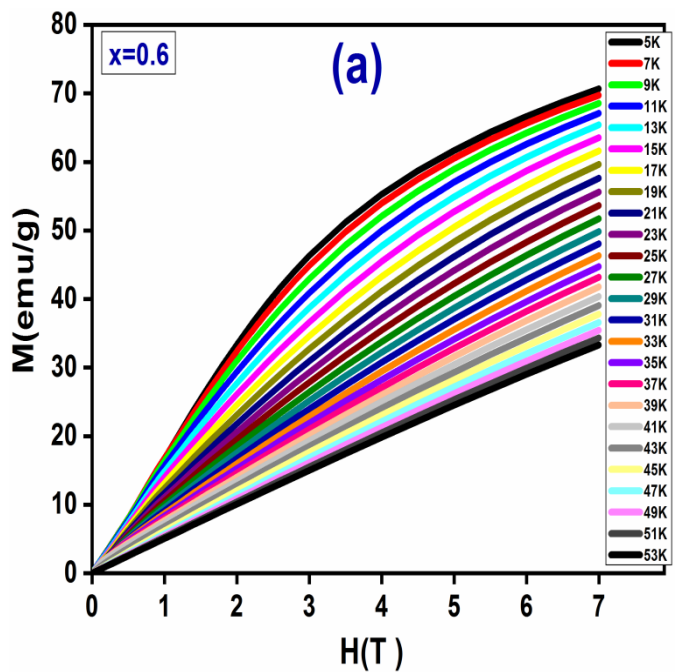
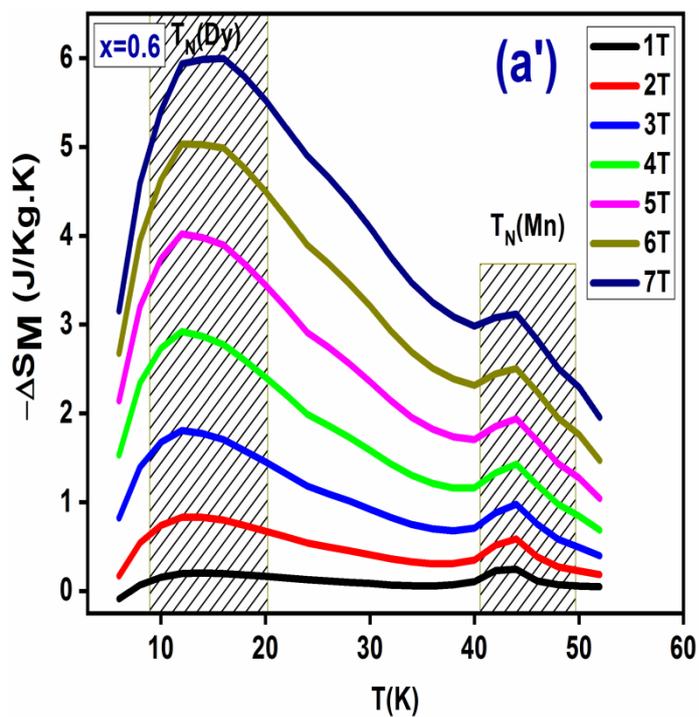
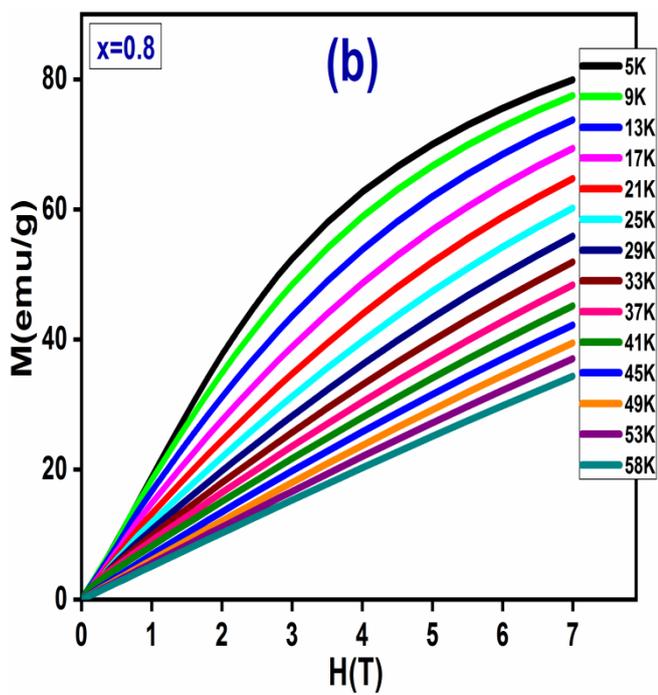
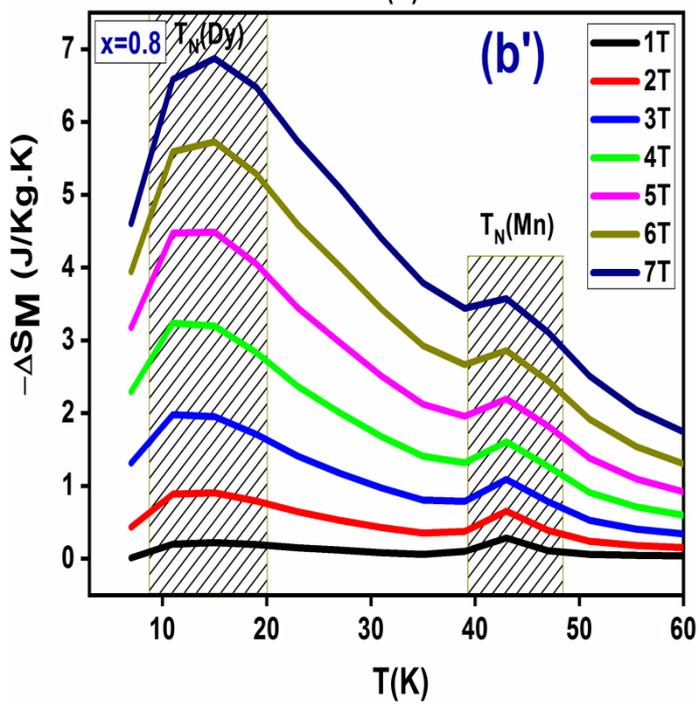

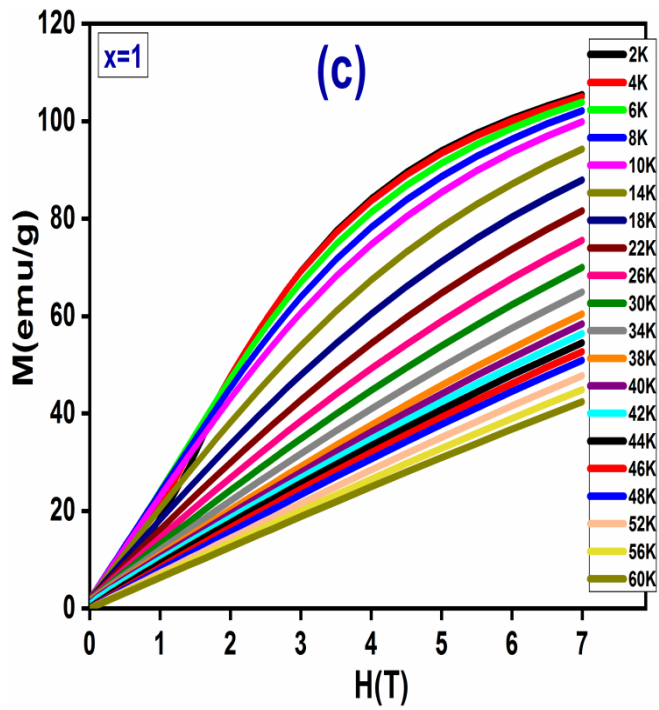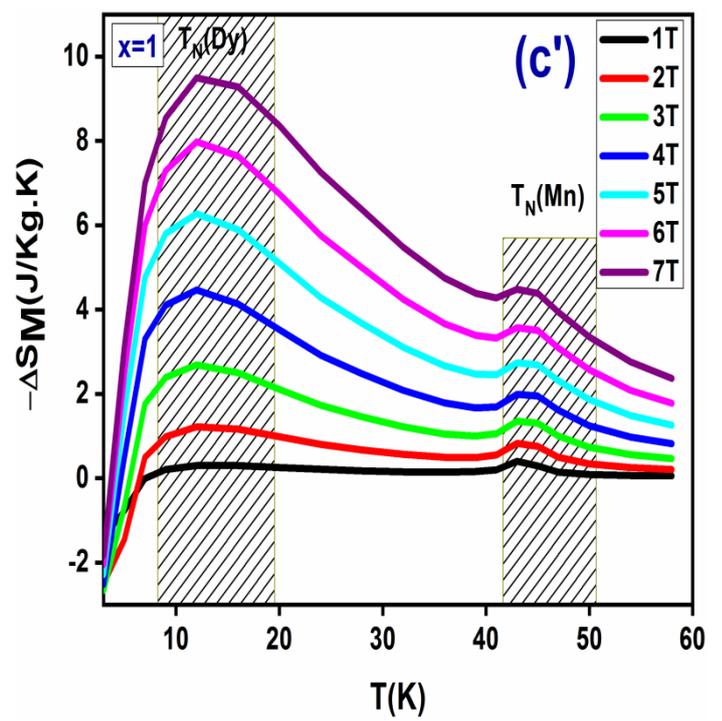

Fig. 6

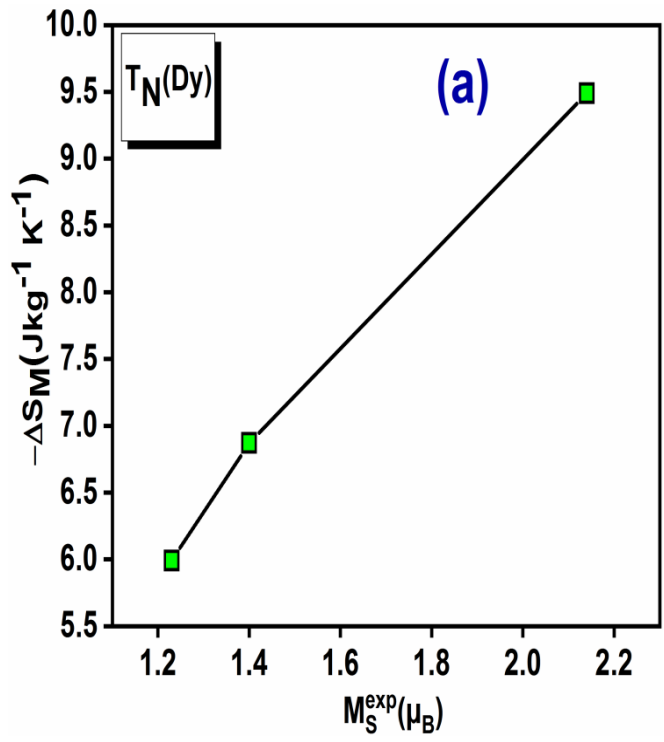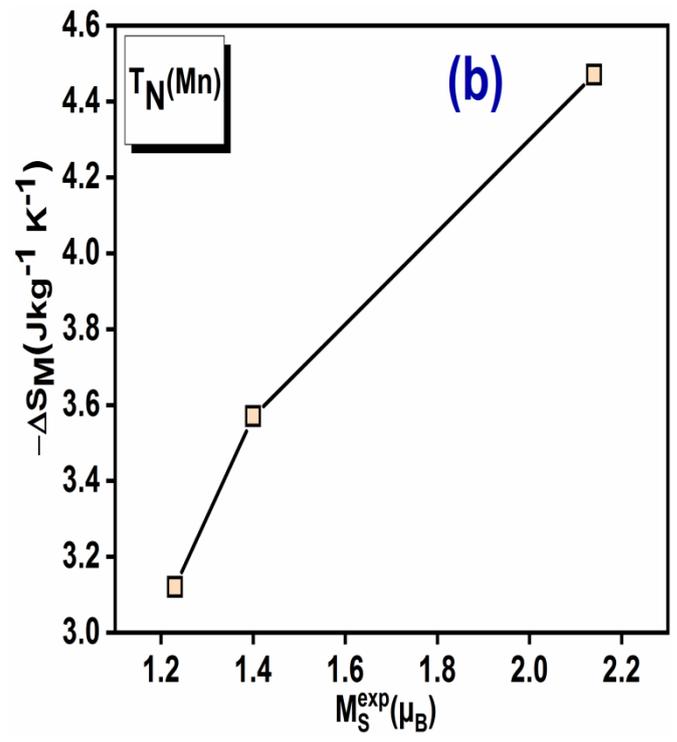

Fig. 7

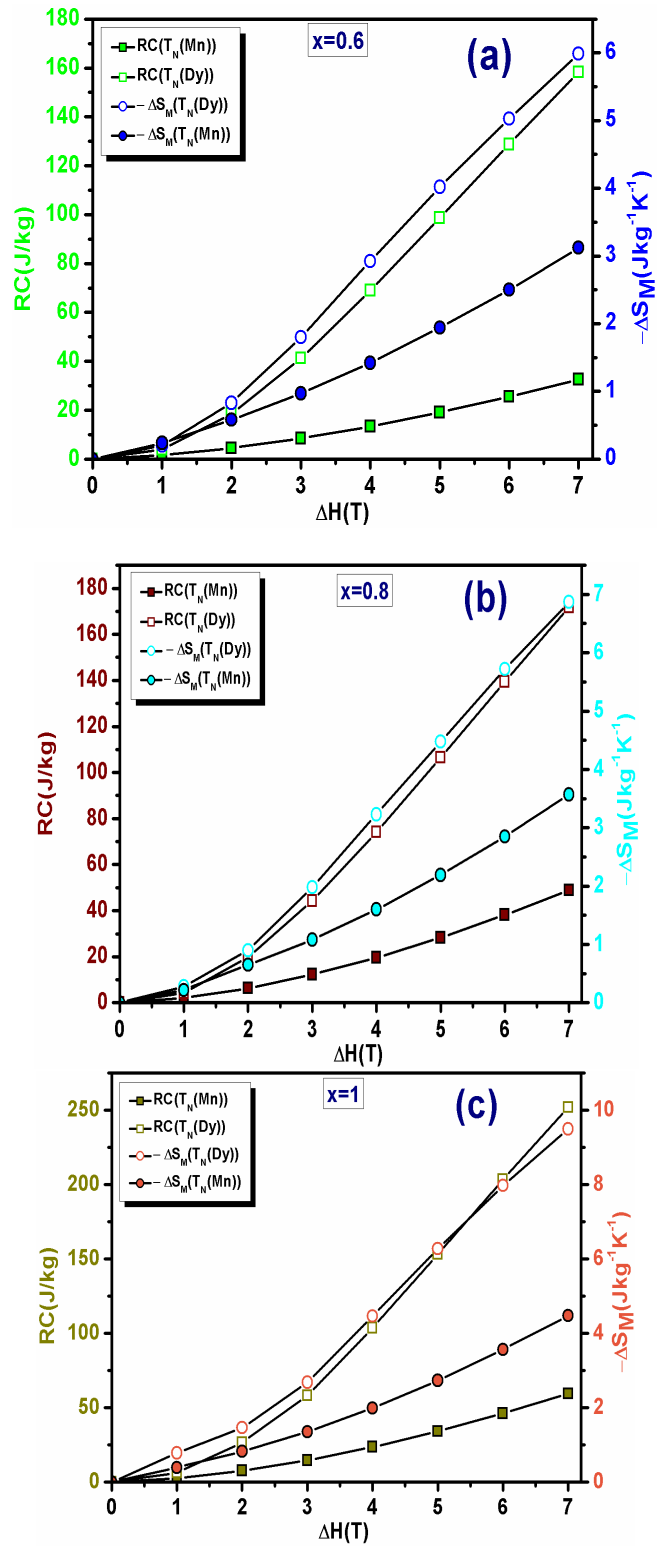

**Fig. 8**

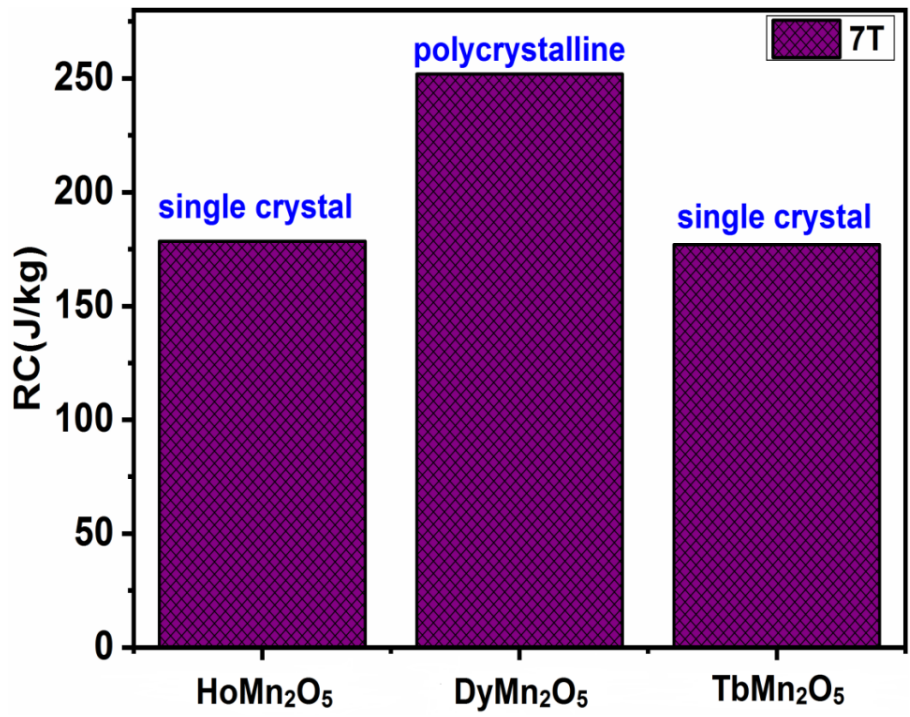

Fig. 9